\documentclass[pra,aps,onecolumn,10pt,superscriptaddress]{revtex4-2}
\usepackage{amsmath,amsfonts,amssymb,graphics,graphicx,epsfig,color,times}
\usepackage[utf8x]{inputenc}
\usepackage{color}
\usepackage{bbm, dsfont} 
\usepackage{subfigure}
\usepackage{mathrsfs}
\usepackage{verbatim}
\usepackage{setspace}
\usepackage{esint}
\usepackage{lineno}
\usepackage[unicode=true]
 {hyperref}
\usepackage{lineno}
\begin{document}
\title{High-fidelity multipartite entanglement creation in non-Hermitian qubits}
\author{C.-Y. Liu}
\email{316497z@gmail.com}
\affiliation{Institute of Atomic and Molecular Sciences, Academia Sinica, Taipei 10617, Taiwan}
\author{C. G. Feyisa}
\affiliation{Institute of Atomic and Molecular Sciences, Academia Sinica, Taipei 10617, Taiwan}
\affiliation{Molecular Science and Technology Program, Taiwan International Graduate Program, Academia Sinica, Taiwan}
\affiliation{Department of Physics, National Central University, Taoyuan 320317, Taiwan}
\author{Muhammad S. Hasan}
\affiliation{Institute of Atomic and Molecular Sciences, Academia Sinica, Taipei 10617, Taiwan}
\author{H. H. Jen}
\email{sappyjen@gmail.com}
\affiliation{Institute of Atomic and Molecular Sciences, Academia Sinica, Taipei 10617, Taiwan}
\affiliation{Physics Division, National Center for Theoretical Sciences, Taipei 10617, Taiwan}

\date{\today}
\renewcommand{\r}{\mathbf{r}}
\newcommand{\f}{\mathbf{f}}
\renewcommand{\k}{\mathbf{k}}
\def\p{\mathbf{p}}
\def\q{\mathbf{q}}
\def\bea{\begin{eqnarray}}
\def\eea{\end{eqnarray}}
\def\ba{\begin{array}}
\def\ea{\end{array}}
\def\bdm{\begin{displaymath}}
\def\edm{\end{displaymath}}
\def\red{\color{red}}
\pacs{}
\begin{abstract}
Non-Hermitian quantum systems showcase many distinct and intriguing features with no Hermitian counterparts. One of them is the exceptional point which marks the $\mathcal{PT}$ (parity and time) symmetry phase transition, where an enhanced spectral sensitivity arises and leads to novel quantum engineering. Here we theoretically study the multipartite entanglement properties in non-Hermitian superconducting qubits, where high-fidelity entangled states can be created under strong driving fields or strong couplings among the qubits. Under an interplay between driving fields, couplings, and non-Hermiticity, we focus on generations of GHZ states or GHZ classes in three and four qubits with all-to-all couplings, which allows a fidelity approaching unity when relatively low non-Hermitian decay rates are considered. This presents an ultimate capability of non-Hermitian qubits to host a genuine and maximal multipartite entanglement. Our results can shed light on novel quantum engineering of multipartite entanglement generations in non-Hermitian qubit systems.  
\end{abstract}
\maketitle
\section{Introduction}

Non-Hermitian physics \cite{Ashida2020, Bergholtz2021} provides an effective description of genuine open quantum systems, which emerges in versatile platforms involving photonic, mechanical, and electrical degrees of freedom \cite{Ashida2020}. These platforms, either classical or quantum mechanical, have shown many advancing and intriguing properties in enhanced sensitivity \cite{Wiersig2014, Hodaei2017}, controllable energy transfer \cite{Doppler2016, Choi2017, Zhang2018}, distinct laser sources \cite{Miri2019}, and novel strongly correlated states of light \cite{Ozawa2019, Abbasi2022}. One essential feature in non-Hermitian system is the exception points (EPs) \cite{Minganti2019, Bergholtz2021}, at which both eigenvalues and eigenstates coalesce. They further mark the $\mathcal{PT}$-symmetry phase transitions \cite{Bender2007, El-Ganainy2018, Ozdemir2019, Li2019}, where real energy spectra can be formed and restore the Hermiticity required in Hermitian physics. Essentially, EPs in one-dimensional or high-dimensional systems are relevant for the emergence of non-Hermitian skin effect \cite{Lee2016, Yao2018, MartinezAlvarez2018, Kunst2018, Lee2019, Borgnia2020, Okuma2020, Zhang2020_skin, Zhang2022_skin, Wang2022_skin}, where an extensive number of localized eigenstates are occupied at the system's open boundaries \cite{Wang2024_skin, Jen2024}. 

Intuitively, non-Hermitian phenomena can be modeled using gain and loss introduced to the systems \cite{Miri2019}, where EPs emerge, and they are highly related to and can further be tailored by these nonconservative gains and losses. Recently, maximally entangled states can be generated in non-Hermitian qubits at a shorter timescale when the qubits are weakly coupled utilizing high-order EPs \cite{Li2023, Feyisa2024}. Stable entanglement dynamics has been studied when significant non-Hermiticity is considered \cite{Zhang2024}, while the robustness of tripartite entangled states is shown to be resilient to non-uniform couplings or off-resonant driving fields \cite{Feyisa2024_2}. Meanwhile, the strong driving or the strong coupling regimes are less explored, which should provide new routes to create multipartite entangled states with potentially better performances.    

Here we theoretically study the creation of high-fidelity and stable GHZ states \cite{GHZ1989} or GHZ class through a one-step entanglement process, which is equivalent to a three-qubit entangling gate \cite{Neeley2010}. In this circuit quantum electrodynamics architecture \cite{Wallraff2004, Blais2004, Neeley2010, Dicarlo2010}, we use non-Hermitian qubit platforms with all-to-all couplings \cite{Roy2020, Zhou2023, Wu2024} to explore the possibility to generate GHZ states under strong driving fields or strong couplings. We find that an almost unit fidelity can be achieved when low non-Hermiticity is considered in three-qubit and four-qubit cases. Our scheme, similar to one-step GHZ state creation process \cite{Neeley2010}, can reduce the time and complexity needed to generate large-scale entangled states without requiring sequential and lower-dimension entangling gates. In contrast to low cost of weak couplings in accelerating entanglement generation in non-Hermitian qubits \cite{Li2023, Feyisa2024}, our results promise an alternative scheme to create high-fidelity quantum states useful for applications in quantum computation and quantum information processing. 

The paper is organized as follows. In section 2, we introduce the model of non-Hermitian qubits system with all-to-all couplings. In section 3, we study the GHZ states and GHZ class generations by analyzing the three-tangle dynamics \cite{Coffman2000}, which genuinely describes the properties of multipartite entanglement. In section 4, we extend our analysis to four-qubit system and confirm the creation of GHZ states of four qubits by identifying respective entanglement entropy among the qubits. We discuss and conclude in section 5.

\section{Theoretical Model}

\begin{figure}[t]
	\centering{}\includegraphics[width=0.45\textwidth]{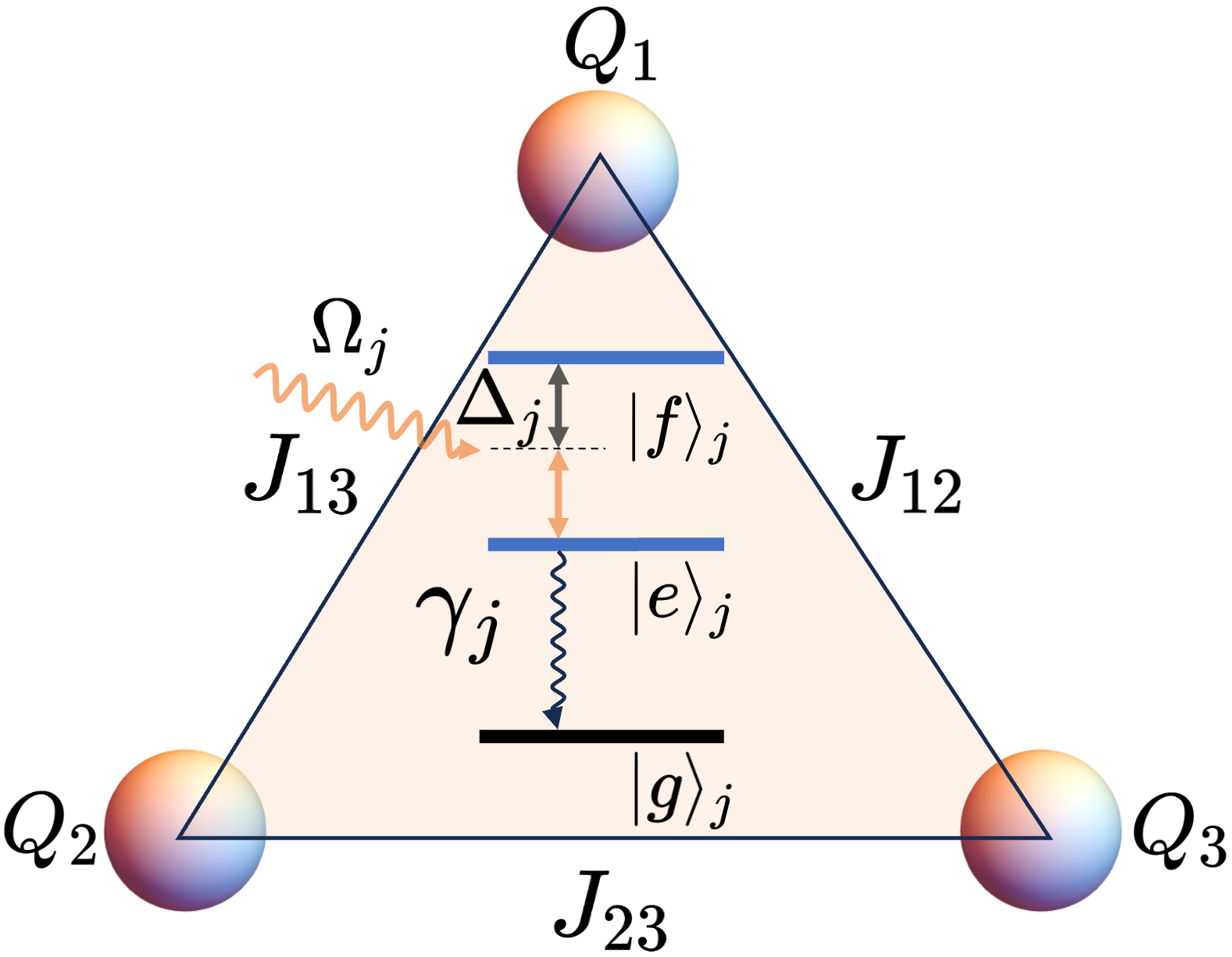}
	\caption{Schematic of three qubits with all-to-all coupling system. Qubit $Q_{1}$, $Q_{2}$, $Q_{3}$ encoded are interconnected with each other with coupling strength $J_{jk}$, while $j,k$ are the qubit number indices. The middle inset depicts the energy configurations of $Q_{j}$, with the driving field $\Omega_{j}$ pumps the qubits states between transitions $\lvert e \rangle_{j} \longleftrightarrow \lvert f \rangle_{j}$ under a detuning $\Delta_{j}$. The decay from the qubit state $\lvert e \rangle_{j}$ to the ground state $ \lvert g \rangle_{j}$ is denoted as $\gamma_{j}$. This effectively reduces the three-level qubit structure to the two-level non-Hermitian qubit systems.}\label{fig1}
\end{figure}

We first consider non-Hermitian qubits interconnecting with each other in a symmetric way \cite{Roy2020} as realized in transmon superconducting qubits \cite{Naghiloo2019}. The Hamiltonian $\hat{H}$ can be written as \cite{Li2023, Feyisa2024},   
\bea
	\hat{H}=\sum^{n}_{j=1}\bigg[(\Delta_j-\frac{i\gamma_j}{2})\hat{\sigma}_{j}\hat{\sigma}^{\dagger}_{j}+\Omega_{j}\hat{\sigma}^{x}_{j}\bigg]\nonumber+\sum^{n}_{j=1}\sum^{n}_{k=j+1}J_{jk}\big(\hat{\sigma}^{\dagger}_{j}\hat{\sigma}_{k}+\hat{\sigma}_{j}\hat{\sigma}^{\dagger}_{k}\big), 
	\label{eq:Hamiltonian}
\eea
where non-Hermiticity is introduced as in the term with decay constants $\gamma_j$, as shown in Fig. \ref{fig1}. All qubits are interconnected with each other \cite{Roy2020, Zhou2023, Wu2024} through a coupling strength denoted by \( J_{jk} \), where \( j \) and \( k \) represent the respective qubit indices. A driving field \( \Omega_{j} \) drives the transitions between the states \( \lvert e \rangle_{j} \) and \( \lvert f \rangle_{j} \) with a detuning denoted as $\Delta_{j}$, and $\hat\sigma_j\equiv|e\rangle_j\langle f|$ with $(\hat\sigma_j^\dag)^\dag=\hat\sigma_j$. The decay rate from the state \( \lvert e_{j} \rangle \) to \( \lvert g_{j} \rangle \) is represented by \( \gamma_{j} \), which is more prominent than the decay from the level $|f\rangle_j$. The dominating \( \gamma_{j} \) effectively reduces the three-level structure in Fig. \ref{fig1} to a two-level configuration and leads to a non-Hermitian model with pure losses. The state $|g\rangle$, on the other hand, can be used as a precursor to monitor the lossy process from the state $|e\rangle$ if it is populated \cite{Li2023, Naghiloo2019}. 

A schematic of non-Hermitian three-qubit systems can be shown in Fig. \ref{fig1}, where $n=3$ is chosen in Eq. (\ref{eq:Hamiltonian}). Throughout the paper, we consider the resonant case where $\Delta_{j} = 0$ and assume a symmetric system with $J_{jk} = J$, $\Omega_{j} = \Omega$, and $\gamma_{j} = \gamma$. This symmetry results in identical quantum dynamics for all qubits, and for convenience we use $J$ as the scaling unit as it is more relevant to investigate the time dynamics of generated high-fidelity entanglement. 

To investigate the entanglement properties of non-Hermitian qubits, we adopt the entanglement entropy $S_{j}$ \cite{Horodecki2009}, three tangle $\tau$ \cite{Coffman2000}, and GHZ state fidelity $F$. Starting from an initial state denoted as $\lvert \psi(t=0) \rangle$, the density matrix $\rho(t)$ and the system dynamics under post-selection \cite{Naghiloo2019} becomes 
\begin{eqnarray}
	\rho\left(t\right) = \lvert\psi\left(t\right)\rangle\langle\psi\left(t\right)\lvert / \lvert \langle \psi\left(t\right) \lvert \psi\left(t\right)  \rangle \lvert, 
	\label{eq:densitynor}
\end{eqnarray} 
with 
\begin{eqnarray}
	\lvert\psi\left(t\right)\rangle =  e^{-i \hat{H} t}\lvert\psi(t=0)\rangle, 
	\label{eq:Propagatorstate}
\end{eqnarray}
and the normalized wave vector $\lvert \widetilde{\psi} \left(t\right) \rangle$ can be expressed in terms of $2^3=8$ orthogonal bare states,  
\bea
	\lvert \widetilde{\psi} \left(t\right) \rangle &=& \frac{\lvert \psi\left(t\right) \rangle}{\sqrt{\lvert \langle \psi\left(t\right) \lvert \psi\left(t\right) \rangle \lvert}},\nonumber \\
	&=& \sum_{i,j,k=e,f} a_{ijk}(t) \lvert ijk \rangle.
	\label{eq:Propagatorstatenor}
\eea
It is noted that $H\neq H^{\dagger}$, which leads to the unnormalized wave vector under the system evolution associated with the non-Hermiticity in non-Hermitian models. The von Neumann entanglement entropy $S_{j}$ can be calculated as \cite{Islam2015}
\begin{eqnarray}
	\mathcal{S}_{j}=-{\rm Tr}[\hat\rho_{j}(t){\rm ln}\hat\rho_{j}(t)],
	\label{eq:von Neumann entropy}
\end{eqnarray} 
with $\rho_{j}$ the reduced density matrix of the $j$-th qubit, obtained by tracing out all other Hilbert space components. For symmetric system considered here, we note that $S_{j} = S$ in general. 

The three tangle $\tau$ can be calculated by the formula \cite{Coffman2000}
\begin{eqnarray}
	\tau=4|d_{1}-2d_{2}+4d_{3}|,\label{e4}
	\label{eq:three tangle}
\end{eqnarray}
where 
\bea
	d_1 &=& \left(a_{eee} a_{fff}\right)^2 + \left(a_{eef} a_{ffe}\right)^2 + \left(a_{efe} a_{fef}\right)^2 + \left(a_{fee} a_{eff}\right)^2, \\
	d_2 &=& a_{eee} a_{fff} \big(a_{eff} a_{fee} + a_{fef} a_{efe} + a_{ffe} a_{eef}\big)+ a_{eff} a_{fee} a_{fef} a_{efe} \nonumber\\
	&+& a_{eff} a_{fee} a_{ffe} a_{eef}+ a_{fef} a_{efe} a_{ffe} a_{eef}, \\
	d_3 &=& a_{eee} a_{ffe} a_{fef} a_{eff} + a_{fff} a_{eef} a_{efe} a_{fee},
\eea
which genuinely quantifies the maximally entangled state in the GHZ class $\tau=1$, while excludes any biseparable states or W states as $\tau=0$. This is also called residual entanglement that is genuinely distributed among all three qubits \cite{Coffman2000}. Meanwhile, the fidelity $F$ provides a straightforward measurement of the likelihood to the target state $\lvert \phi \rangle$, which reads 
\bea
	F = \lvert \langle \tilde \psi \left( t \right)\lvert \phi \rangle \lvert^{2}. 
\eea

In this work, we focus on the properties of three and four qubits systems, where we target the GHZ states $ \lvert \phi \rangle$ $=$ $\frac{1}{\sqrt{2}}( \lvert eee \rangle$ $+$ $\lvert fff \rangle)$ \cite{Dur2000} or GHZ class for three qubits as an example. This state yields the maximally entanglement distribution among three qubits but with vanishing biseparable concurrences \cite{Wooters1998}. This leads to $\tau = 1$ when perfect GHZ state is created. Furthermore, under LOCC (Local unitary operations and classical communication), the states belonging to GHZ class are in fact equivalent to the GHZ state, given that under LOCC, the entanglement properties will not change \cite{Nielsen 2001}. 

\begin{figure}[t]
	\centering{}\includegraphics[width=0.7\textwidth]{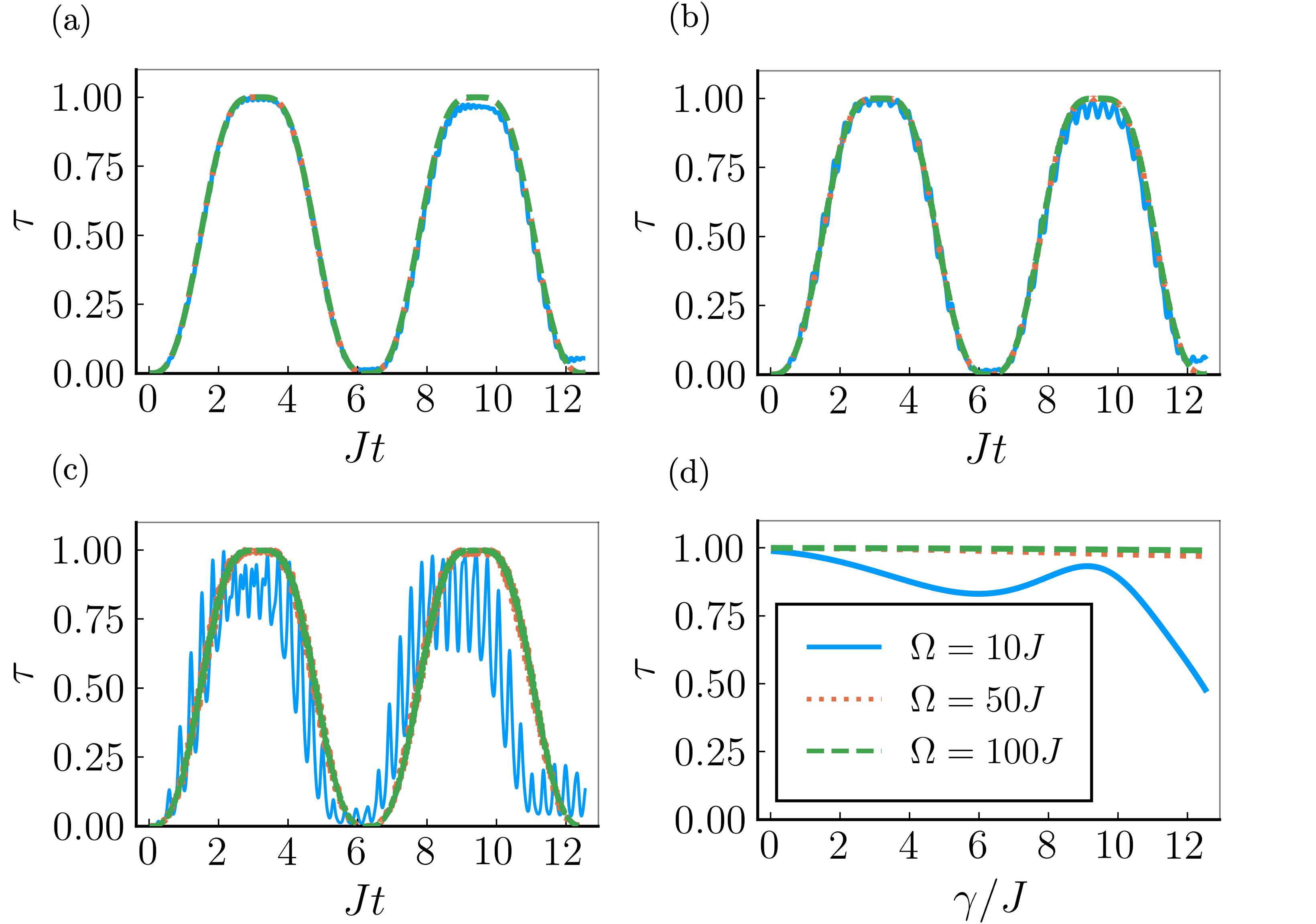}
	\caption{Three tangle dynamics in the strong driving regime. We demonstrate three examples of $\Omega = 10J$ (blue solid line), $\Omega = 50J$ (orange dotted line), and $\Omega = 100J$ (green dashed line) in all plots, where three tangle $\tau$ dynamics versus a dimensionless time $Jt$ at (a) vanishing $\gamma$, (b) $\gamma = 1J$, and (c) $\gamma = 6J$. (d) Three tangle properties for different $\gamma/J$ at $Jt = \pi$.}\label{fig2}
\end{figure}

\section{GHZ states in three qubits} 

Here we study the GHZ states generations in three non-Hermitian qubits system under strong driving fields and strong coupling regimes, respectively. The former drives individual qubits uniformly, while the latter exchanges spin excitations between any pair of qubits in an all-to-all fashion.  

\subsection{Strong driving regime} \label{sec:strong_driving}

Under a strong driving regime when $\Omega \gg J$, $\gamma$, in Fig. \ref{fig2} we calculate the three tangle $\tau$ from Eq.~\ref{eq:three tangle} given an initial pure state as $\lvert \psi \rangle = \left(\lvert f\rangle \right)^{\otimes3}$.  We note that a different initial state of $\lvert \psi \rangle = \left(\lvert e\rangle - i \lvert g\rangle \right)^{\otimes3}$ can also show similar dynamics as in Fig~\ref{fig2}. In Figs. \ref{fig2}(a), \ref{fig2}(b), and \ref{fig2}(c), we compare different time dynamics of $\tau$ when we increase the dissipation rates $\gamma$. They oscillate at a period of $Jt=2\pi$ and can achieve quite high three tangle close to unity at a half period. However, when a relatively comparable $\gamma=6J$ under a moderate driving field $\Omega$ is considered in Fig. \ref{fig2}(c), $\tau$ becomes mitigated and deviates significantly away from the strong coupling regime. 

To illustrate the robustness of $\tau$ against the decay rate $\gamma$, in Fig. \ref{fig2}(d) we plot three tangle at a chosen time $Jt = \pi$ when maximal $\tau$ emerges. When significant $\Omega$ is applied, $\tau$ can sustain around unity over a wide range of finite $\gamma$, which indicates the robustness of states in the GHZ class. We have also checked that the entanglement entropy $S$ is close to $\log2$ corresponding to the maximal $\tau$, which reaffirms the creation of maximally entanglement in the GHZ classes. The fidelity $F$ is calculated to be $0.9999$ and $0.999$ for $\gamma =0$ and $6J$, respectively, when $\Omega=100J$ is considered. Therefore, the system under strong driving regime can provide significant improvement in entanglement generations as well as the fidelity, but at a cost of demanding resource of driving fields. This is in contrast to a weak coupling regime under a moderate driving field, where the state fidelity is not able to be maximized even though a faster generation of entangled states can be manifested \cite{Li2023, Feyisa2024}. 

As a final remark, a revival of $\tau$ in Fig. \ref{fig2}(d) at a moderate $\Omega=10J$ further displays the capability to optimize the multipartite entanglement property even at a moderate decay rate $\gamma\approx 9J$. This as well promises a future design of optimized system parameters $(\Omega,\gamma)$ which lead to good enough $\tau$ and associated fidelity without requiring costly resources of driving fields.   

\begin{figure}[t]
	\centering{}\includegraphics[width=0.7\textwidth]{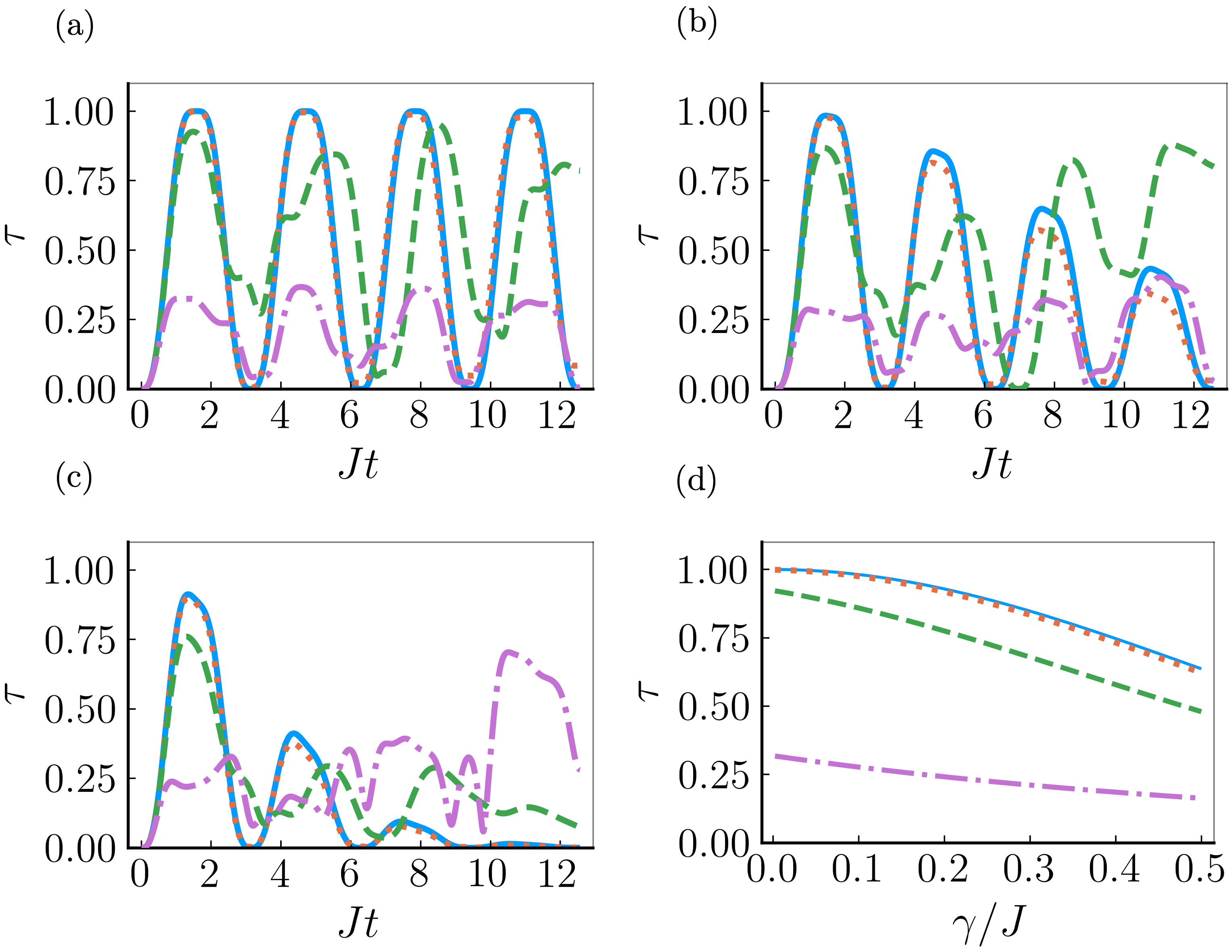}
	\caption{Three tangle dynamics under the strong coupling regime from an initial spin coherent state $\lvert \psi \rangle =   \left( \lvert f \rangle + e^{i \phi} \lvert e \rangle \right)^{\otimes 3}$ with $\phi\approx 0.288\pi$. We illustrate several cases of $\Omega = 0$ (blue solid line), $\Omega = 0.01J$ (orange dotted line), $\Omega = 0.1J$ (green dashed line), and $\Omega = 0.5J$ (purple dash-dot line), respectively, at (a) $\gamma = 0$, (b) $\gamma = 0.1J$, (c) $\gamma = 0.25J$. (d) Three tangle versus $\gamma/J$ for different driving fields when $Jt =\pi/2$. We note that different $\phi$'s show similar time dynamics and properties in our plots.}\label{fig3}
\end{figure}

\subsection{Strong coupling regime} \label{sec:strong_coupling}

Next we investigate the strong coupling regime when $J$ dominates the other system parameters, that is $J \gg \Omega, \gamma$. This regime corresponds to the $\mathcal{PT}$ symmetric phase as it yields real eigenvalues. We specifically choose the initial states as the spin coherent states $\lvert \psi \rangle =   \left( \lvert f \rangle + e^{i \phi} \lvert e \rangle \right)^{\otimes 3}$ to achieve high-fidelity GHZ class in such regime, with $\phi $ being an arbitrary phase.

In Fig. \ref{fig3}, we again plot three tangle dynamics $\tau$ similar to Fig. \ref{fig2}. We find that the maximal $\tau$ emerges at $Jt =(2m-1)\pi/2$ with an integer $m$ in Figs. \ref{fig3}(a) and \ref{fig3}(b) when $\gamma$ or $\Omega$ is relatively smaller than $J$. The robustness of three tangle can be seen in Fig.~\ref{fig3}(d) when $\gamma$ is made much smaller than $J$. Although a finite $\Omega$ does not benefit too much on the system's three tangle at a short time, but at a longer time, $\tau$ with a larger driving field can surpass the results with lesser $\Omega$, as shown in Fig. \ref{fig3}(c). This intricate interplay between driving fields and qubit couplings showcases the essential feature to control and tailor the properties of multipartite entanglement in non-Hermitian qubits. 

\section{Four qubtis}

\begin{figure}[!t]
	\centering{}\includegraphics[width=0.75\textwidth]{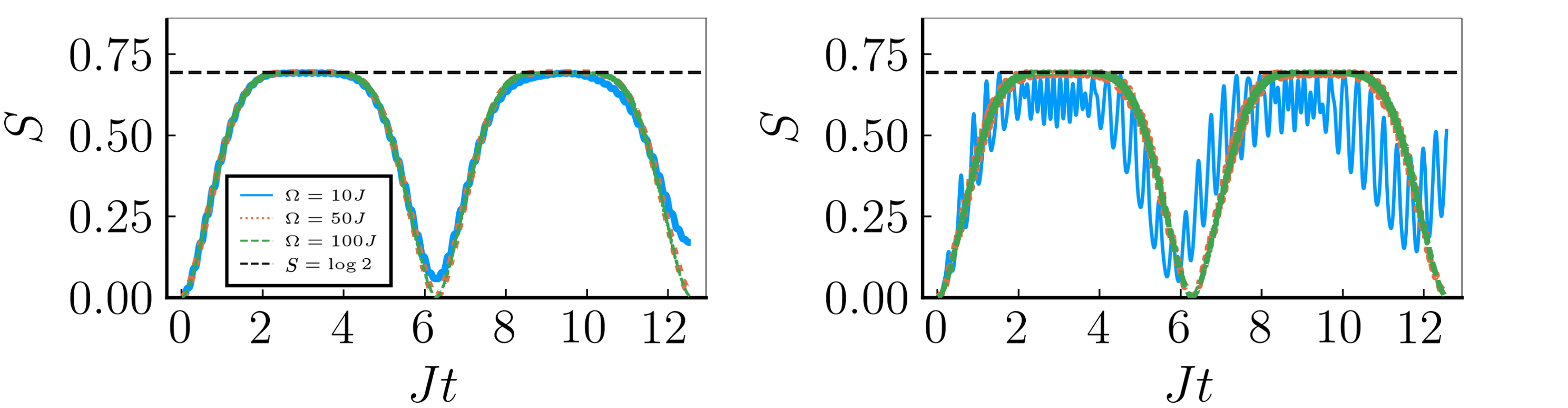}
	\caption{Entanglement entropy $S$ dynamics under the strong coupling regime for four qubits. We plot the $S$ versus dimensionless time $Jt$ when (a) $\gamma = 0$J and (b) $\gamma=6J$ for different driving fields $\Omega$ as in Fig. \ref{fig2}. The system generates a particular GHZ class state during the period when $S$ reaches around $\log2$.}\label{fig4}
\end{figure}

Finally, we extend our analyses to four-qubit systems in Fig. \ref{fig4}. Similarly in three qubits, the dynamics yields interesting results with periodically entanglement entropy $S \approx \log 2$ under the strong driving regime. The entanglement creation shows stable oscillations with broad plateaus near $\log 2$ at a period of $Jt=2\pi$. Similar to previous conclusions on the role of $\gamma$, the entanglement entropy degrades as $\gamma$ increases along with fast oscillations. Meanwhile, the robustness of $S$ can be maintained as long as $\Omega$ is made significant. 

At the time $Jt=\pi$ when $S$ is maximized for $\Omega=100J$, we compare the state preparations with GHZ state $\lvert \psi_{\text{GHZ}} \rangle$ $=$ $(\lvert eeee \rangle + \lvert ffff \rangle)/\sqrt{2}$. The generated state $\lvert \psi \left(t\right) \rangle$ indeed resembles the $\lvert \psi_{\text{GHz}} \rangle$ only up to single-qubit rotations, which is $ Z\left(\frac{3\pi}{4}\right)\otimes I\otimes I \otimes I) \lvert \psi\left(t\right)\rangle \approx \lvert \psi_{\text{GHz}}\rangle$ with fidelity $F \approx 0.998$ and $0.99$ for $\gamma = 0$ and $6J$, respectively. The single-qubit phase gate $Z\left(\phi\right)$ \cite{Nielsen 2001} can be denoted as 
\bea
Z\left(\phi\right)\equiv
\begin{bmatrix}
	e^{i \phi} & 0 \\
	0 & 1
\end{bmatrix}.
\eea

\section{Conclusion and discussion} 

In conclusion, we theoretically explore the genuine GHZ state creations in three and four qubits in non-Hermitian transmon superconducting platforms. With symmetric all-to-all couplings, we focus on the strong driving and the strong coupling regimes. Via calculations of three tangle in three qubits and entanglement entropy characterizations in four qubits, we are able to identify the parameter regimes where robust and genuine multipartite entanglement generations emerge. Our results suggest that high-fidelity state preparations in the GHZ class can be achieved with a fidelity surpassing $0.999$. Our scheme resembles a one-step GHZ gate operation without seeking for quantum circuits by combing multiple two-qubit gates \cite{Gu2021}. This saves more time for entangling operation, and our results further promise optimal conditions with only a moderate driving field to accomplish favorable entanglement as well as associated fidelity. Non-Hermitian qubits system thus can offer great potential for applications in quantum information processing and quantum computation \cite{Li2023,Zhang2024}. 

\section*{ACKNOWLEDGMENTS}
We acknowledge support from the National Science and Technology Council (NSTC), Taiwan, under the Grants No. 112-2112-M-001-079-MY3 and No. NSTC-112-2119-M-001-007, and from Academia Sinica under Grant AS-CDA-113-M04. We are also grateful for support from TG 1.2 of NCTS at NTU.

\end{document}